V.A. Maisheev, Y.A. Chesnokov, P.N. Chirkov, I.A. Yazynin
*IHEP, Protvino, RF*

D. Bolognini, S. Hasan, M. Prest
*Università dell'Insurbia, Milano, Italy*

E. Vallazza
*INFN, Trieste, Italy*

# Channeling Radiation of Positrons with Energy in the Region of 100 GeV in Single Crystals

**Abstract**

The process of radiation of 120-GeV positrons moving in a channeling regime in (011) plane of a single crystal was considered. At the beginning on the basis of the theory of nonlinear oscillations the trajectory of moving positrons at different initial conditions were derived. Then taking into account the nonlinearity of motion the distribution function over oscillation amplitudes of channeling particles was found. After this the intensity of radiation at different initial conditions was calculated with the help of two various methods. These results may be useful for comparison with experimental data at positron energies from 100 and more GeV.

# Introduction

By this time the considerable number of experimental and theoretical works is devoted to researching the radiation at plane channeling of high energy positrons in monocrystals (see [1, 2, 3] and the literature quoted there). This radiation arises during the motion of a charged particle under a small angle in relation to a crystallographic plane and (in a case of ultrarelativistic positrons with energies up to ~20 GeV) is monochromatic enough and is characterized by high intensity. At energies of positrons more than ~20 GeV monochromaticity of the radiation strongly degrades. The majority of experiments on study the radiation during channeling of the relativistic positrons was executed at energies from several hundreds of MeV to tens GeV where the distinct peak in intensity of radiation dominates which is gradually smeared at energies more than ~20 GeV [3] due to increase of the radiation intensity of high harmonics. Concrete theoretical consideration of the process is also concentrated basically to the specified range of energies. A large number of interesting data has been received during the experiment [4] where the radiation of leptons at energy of 150 GeV in the straight crystals was investigated.

In September, 2009 in CERN the experiment INSURAD devoted to research of radiation at various orientations of bent monocrystals has been made at energy of positrons of 120 GeV. In particular, the big statistics has been received on radiation of positrons during their motion in plane fields of bent silicon monocrystals which is processed now. Continuation of this experiment is planned in 2010, in particular, it is supposed to measure radiating processes in straight (not bent) monocrystals for the purpose of correct comparison of all received data. It is powerful argument in favour of theoretical research of the specified processes (and in particular radiations during plane channeling in the straight and bent monocrystals) at energies more than 100 GeV.

Thus, the given paper is devoted to theoretical consideration of radiation process of the ultrarelativistic positrons with energy of an order of 100 GeV, channeling in straight crystals.

Radiation characteristics of relativistic particles with the set energy $E$ are defined by its motion in a monocrystal electric field. Therefore, first of all, it is necessary to describe this motion. For this purpose it is necessary to know the electric field or the potential distribution in crystallographic planes. Such field (potential) is found analytically on the basis of relations for an atom field in the model of Moliere or, more precisely, on the basis of approximation of the corresponding data received as a result of x-ray measurements. Instead of exact representation of an electric field (or potential) their simplifications (so-called modeling potentials) are often used by means of rather simple functions. The lack of such consideration is that it is impossible to describe the plane fields well enough by some simple functions in the consent with calculations using the Moliere model or any other realistic model of atom. Another approach is used in paper [5] where the field was represented as Fourier series, and it has been shown that plane potentials of a field of monocrystals can be described by polynomials of rather high degree which practically (an error nearby 1 % for a polynomial of 14 degree) do not differ from initial exact representations.

It is obvious that a potential in the central part between crystal planes has the parabolic form, but near to the planes it is strongly nonlinear, because it should provide a zero electric field on the crystal planes due to similarity of plane cells and due to physical continuity. Besides, it is difficult to find an exact solution of the motion equations with the help of asymptotic methods of nonlinear oscillations [6] (and used in [5, 7]) for the case of channeled particles making periodic motion in such nonlinear interplanar potential well with the big amplitudes and closed phase trajectories which are close to a separatrix, separating the channeled particles from the over-barrier ones. Moreover, even insignificant perturbations, such as irregular displacement of planes, will lead to formation of the so-called homoclinic structures on branches of the separatrix [9÷13] and to formation of a stochastic layer nearby the separatrix. In this case the extraordinary complexity of motion in the vicinity of separatrix was known still to H. Poincaré [12÷14]: "… Complexity of this picture of motion so amazes that I do not try to represent it at all". Till now it was not possible to receive a strict estimation of width of the stochastic layer, and the results presented usually are based on approximate description of motion in the vicinity of separatrix in presence of perturbations. Therefore in this paper we have tried as much as possible to do without searching the exact solution of the equations of motion in such nonlinear potential (without perturbations) for all channeled particles, and limited ourselves by using of the Hamiltonian formalism for the description of the motion and for finding the characteristic parameters (or, to be more exact, the functions) of the motion. Among them such functions as dependence of frequency and multipole parameter $\rho$ from amplitude of periodic motion, amplitude distribution of the channeled positrons. Use of such formalism has allowed to overcome some difficulties in determining the radiation occurring at sufficiently large amplitudes (see [5, 7]).

The radiation type of a relativistic particle depends on the value of multipole parameter $\rho$. When $\rho \ll 1$ it corresponds to the interference type (dipole approximation) of the radiation formed along sufficiently large length of the crystal. The case with $\rho \gg 1$ is close to the synchrotron radiation. At $\rho \sim 1$ the intermediate case takes place. Analyzing some radiation process it is necessary to consider the whole ensemble of the channeled particles in the corresponding phase space because the motion parameters (amplitude of oscillations, for example) in a not thick monocrystal are defined by the initial conditions on its input. And as it will be shown, various types of radiation generated by particles with different initial conditions can be realized in the considered process.

The parameter $\rho$ becomes an order of 1 at planar channeling of positrons (at least for an appreciable part of the particles) starting from energies of several GeV. At energies of tens GeV a major portion of the positrons is characterized by parameter $\rho$ from 1 to several units. In this case the calculations should take into account the nondipole character of radiation. The corresponding mathematical apparatus for the radiation during a periodic motion can be found in the monograph [1] (quantum and classical cases in vacuum) and paper [8] for the transparent medium (a classical case). At energies of positrons 100 GeV and more the parameter $\rho$ can exceed 20 units for a considerable part of the particles. In this case one can expect the complication of the procedure of calculations given in [1] for planar motion as the great number of harmonics starts to be radiated. On the other hand the authors of [1] declare that in this case the radiation has practically a magnetic bremsstrahlung character and recommend to use the corresponding formulae for the calculations.

In view of everything told above, in the given work we wish to receive the following results:
1. on the basis of the Hamiltonian formalism to consider the motion of an ultrarelativistic particle in the real plane potential of a monocrystal and to study the influence of nonlinearities on an ensemble of particles captured in a mode of channeling;
2. to investigate and compare various methods of calculation the intensity of positron radiation with energies an order of 100 GeV at their different initial conditions on an input in a monocrystal;
3. to give some predictions for the radiation intensity of the channeled positrons which it will be possible to measure in the future experiment INSURAD.

The received results will be a basis for the further consideration of the radiation process in a bent monocrystal.

## §1. Interplanar one–dimensional motion of channeled positrons

The motion of a charged ultrarelativistic particle in the interplanar electric field D of a monocrystal can be described by the following system of equations

$$\frac{E}{c^2}\frac{d^2x}{dt^2} = eD(x) \quad , \quad \frac{d^2y}{dt^2} = 0 \quad , \quad \frac{ds}{dt} = c\left(1 - \frac{1}{2\gamma^2} - \frac{1}{2c^2}\left(\left(\frac{dx}{dt}\right)^2 + \left(\frac{dy}{dt}\right)^2\right)\right).$$

Where: $x, y, s$ - the Cartesian co-ordinates of a particle (the electric field D is directed along the axis $x$); $E, e, \gamma$ - energy, charge and gamma factor of a particle, accordingly; $t$ - time, $c$ - velocity of light. At certain initial conditions these equations describe the particle motion in the mode of channeling. In this case the first equation describes periodic motion along coordinate $x$, while the third equation reflects the influence of transverse motion of a particle on longitudinal one. Despite of a relatively small value, this change of longitudinal velocity of a particle is taken into account in calculations of radiation intensity. From the above equations it is seen that the problem of finding the trajectory of a particle in three-dimensional space is reduced to finding the function $x(t)$.

In this paragraph we will consider periodic (generally nonharmonic) motion of positrons with energy of $E = 120\,\text{GeV}$ in the interplanar potential of a straight crystal Si with orientation (011). We will define the basic characteristics of this motion which are the most essential to a spectrum of radiation of positrons with the specified energy during their passage through a crystal. These characteristics, first of all, are the dependence of frequency, multipole parameter $\rho$ and density of distribution of positrons on the amplitude of motion.

The interplanar potential is calculated for silicon at a room temperature as it is described in work [5]. At first the exact expression for planar electric field of a monocrystal (in this case on the basis of x-ray measurements of atomic form-factors of silicon) in the form of a Fourier series has been received. Then this field was expanded in a series of orthogonal Legendre polynomials, which has allowed us to present it as a polynomial of interplanar coordinate $x$. It is obvious that this method allows to receive the representation of electric potential of a field as much as close to its exact value. We have limited ourselves by its description with an accuracy nearby 1 % that corresponds to a polynomial of 14 degree. Thus, the interplanar potential of interaction of a positron in a straight crystal is defined by expression

$$U(\xi) = -\frac{d}{2} \sum_{k=1}^{7} \frac{\alpha_k}{2k} \xi^{2k} , \qquad (1)$$

where: $\xi = \dfrac{x}{d/2}$ -normalized interplanar coordinate, $\xi \in [-1, +1]$;

$d = 1.92\,\text{Å}$ - interplanar distance in (011) channel;

$\vec{\alpha} = (-32.21 \quad 13.86 \quad -443.78 \quad 2340.52 \quad -5315.05 \quad 4811.79 \quad -1375.13)$ in [eV/ Å];

such values of $\alpha_k$ provide $dU/d\xi = 0$ at $\xi = \pm 1$.

The dependence $U(\xi)$ is shown in the following drawing, where $U_o = 21.873\,\text{eV}$ - level of potential barrier.

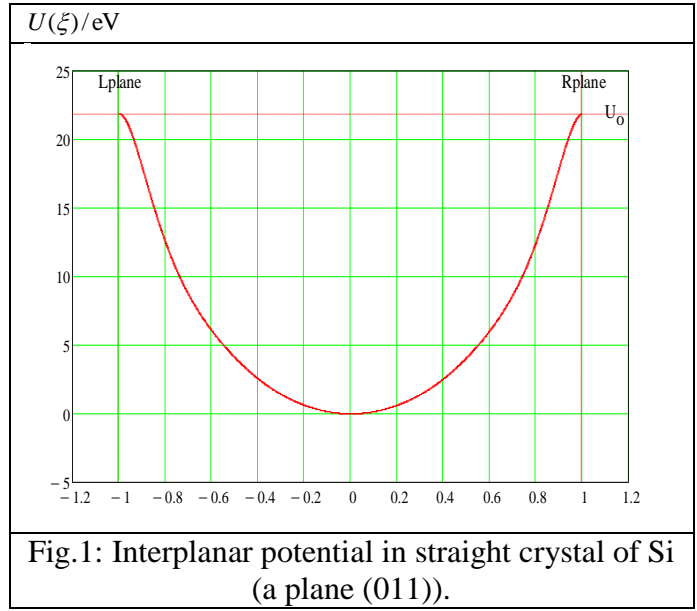

Fig.1: Interplanar potential in straight crystal of Si (a plane (011)).

In a Fig. 2 the dependence of normalized frequency $\Omega(\xi_m)$ on the amplitude of periodic motion is shown

$$\Omega(\xi_m) = \omega(\xi_m)/\omega_o ,$$

where: $\omega_o \equiv \omega(0) = \sqrt{2|\alpha_1|c^2/Ed} \cong 5.013 \times 10^{13}/\text{sec}$ - frequency of oscillations with small (zero) amplitudes in the potential hole $U(\xi)$ shown above. The maximum displacement of periodic motion is interpreted as amplitude $\xi_m$. The technique of getting this dependence was the standard one (see, for example, [9]). Namely, the motion of a positron in normalized potential well $\tilde{U}(\xi)$ defined by the relation

$$\tilde{U}(\xi) = -\frac{1}{|\alpha_1|} \sum_{k=1}^{7} \frac{\alpha_k}{2k} \xi^{2k} \quad ,$$

is described by the canonical equations

$$\frac{d\xi}{d\tau} = p \quad \text{and} \quad \frac{dp}{d\tau} = -\frac{d\tilde{U}(\xi)}{d\xi}, \qquad (2)$$

where: $\tau = \omega_o t$ - the dimensionless time (phase);

$$H(\xi, p) = \frac{p^2}{2} + \tilde{U}(\xi) = \varepsilon \quad \text{- Hamiltonian.} \qquad (3)$$

The Hamiltonian in this case is the integral of motion. To find the frequency (which is reverse to the period) it is convenient to pass from canonical variables $(\xi, p)$ to a new Hamiltonian with canonical variables "angle-action" $(\theta, J)$ with the help of canonical transformation. Using of such new variables is a convenient way to get frequency of a periodic motion, not demanding finding-out of details of the motion itself.

The action $J$ in dependence on transverse "energy" $\varepsilon$ of the motion in a well $\tilde{U}(\xi)$ is defined as the integral along a phase trajectory with fixed $\varepsilon$

$$J(\varepsilon) = \frac{1}{2\pi} \oint p(\xi, \varepsilon) \cdot d\xi = \frac{4}{2\pi} \int_0^{\xi_m(\varepsilon)} \sqrt{2(\varepsilon - \tilde{U}(\xi))} \cdot d\xi \quad ,$$

where $\xi_m(\varepsilon)$ - maximum deviation (amplitude) which is defined from equation

$$\tilde{U}(\xi_m) = \varepsilon. \qquad (4)$$

Owing to the biunique dependence $J(\varepsilon)$ for positrons caught in the interplanar channel, in principle, we know (at least numerically) the inverse function

$$\varepsilon = \varepsilon(J) \quad .$$

Particles with $\varepsilon \in [0, \tilde{U}_o]$ are the channeled ones making a limited motion within the channel, and with $\varepsilon > \tilde{U}_o$ are the over-barrier particles whose motion is not limited over $\xi$. New Hamiltonian where as a canonical momentum the action $J$, which is also the adiabatic invariant, is chosen, for the channeled particles is equal

$$\overline{H}(\theta, J) = \overline{H}(J) = \varepsilon(J) \quad .$$

From this we have:

$$\frac{dJ}{d\tau} = -\frac{\partial \varepsilon(J)}{\partial \theta} = 0 \quad , \quad \frac{d\theta}{d\tau} = \Omega(J) = \frac{\partial \varepsilon(J)}{\partial J} \quad \text{- normalized frequency, i.e.}$$

$$\Omega^{-1}(\xi_m) = \frac{4}{2\pi} \int_0^{\xi_m} d\xi / \sqrt{2(\tilde{U}(\xi_m) - \tilde{U}(\xi))}$$

The dependence of normalized frequency $\Omega$ on amplitude $\xi_m$ for the potential (1) is shown in Fig. 2.

The multipole parameter $\rho$ essentially defines the character of radiation of an oscillating channeled positron and is expressed through parameters of plane periodic motion of a particle as follows [1]:

$$\rho = 2\gamma^2 \langle (\upsilon_x/c)^2 \rangle,$$

where the averaging is taken over the motion period. For the channeled positron with the given amplitude of motion $\xi_m$ we have

$$\rho(\xi_m) = 2\gamma^2 \kappa^2 \langle p^2 \rangle = 2\gamma^2 \kappa^2 \Omega(\xi_m) J(\xi_m) = \gamma^2 \kappa^2 \Omega(\xi_m) \frac{4}{\pi} \int_0^{\xi_m} \sqrt{2(\tilde{U}(\xi_m) - \tilde{U}(\xi))} \, d\xi,$$

where $\kappa = d\omega_o / 2c$. For the potential (1) $\kappa \cong 16.052 \cdot 10^{-6}$ at the frequency $\omega_o$ defined above. In Fig. 3 (the continuous line) the exact dependence of multipole parameter on amplitude $\xi_m$ is shown. Thus, at the given potential both dipole and magnetic bremsstrahlung radiation types can be realized.

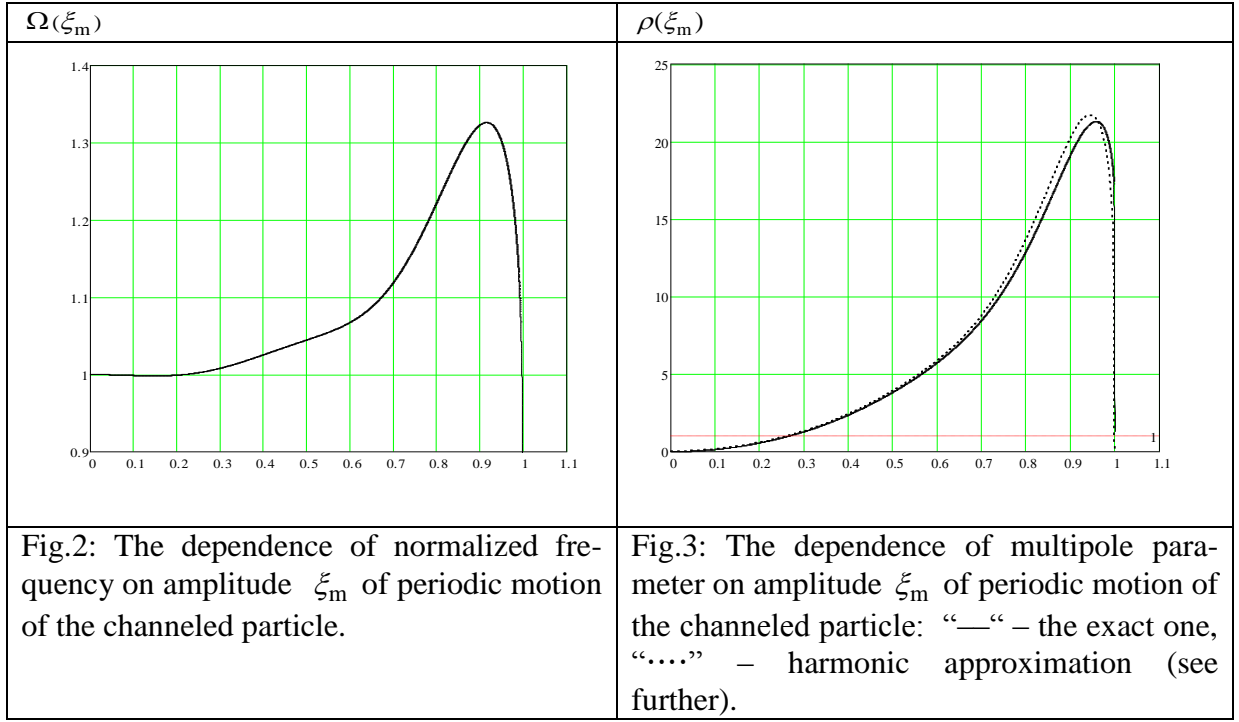

Fig.2: The dependence of normalized frequency on amplitude $\xi_m$ of periodic motion of the channeled particle.

Fig.3: The dependence of multipole parameter on amplitude $\xi_m$ of periodic motion of the channeled particle: "—" – the exact one, "…." – harmonic approximation (see further).

The received frequencies correspond to nonlinear (not harmonic) oscillations. The closer $\xi_m$ to 1, the stronger difference of periodic motion from the harmonic one. Comparison at given $\xi_m$ of the exact numerical decision of the equation of motion with approximating harmonic oscillation with the same $\xi_m$ and normalized frequency $\Omega(\xi_m)$

$$\xi = \xi_m \cos(\Omega(\xi_m) \cdot \tau) \qquad (5)$$

is shown in the Fig. 4a and 4b. In these drawings dependences $\xi$ from $\tau/2\pi$ are seen: continuous lines correspond to the exact numerical decision of the equations of motion, and dotted ones to harmonic approximation. Thus, practically in all range $0 \leq \xi_m \leq 0.980$ we can consider the motion of channeled positrons to be the harmonic one with the calculated values of frequencies $\Omega(\xi_m)$.

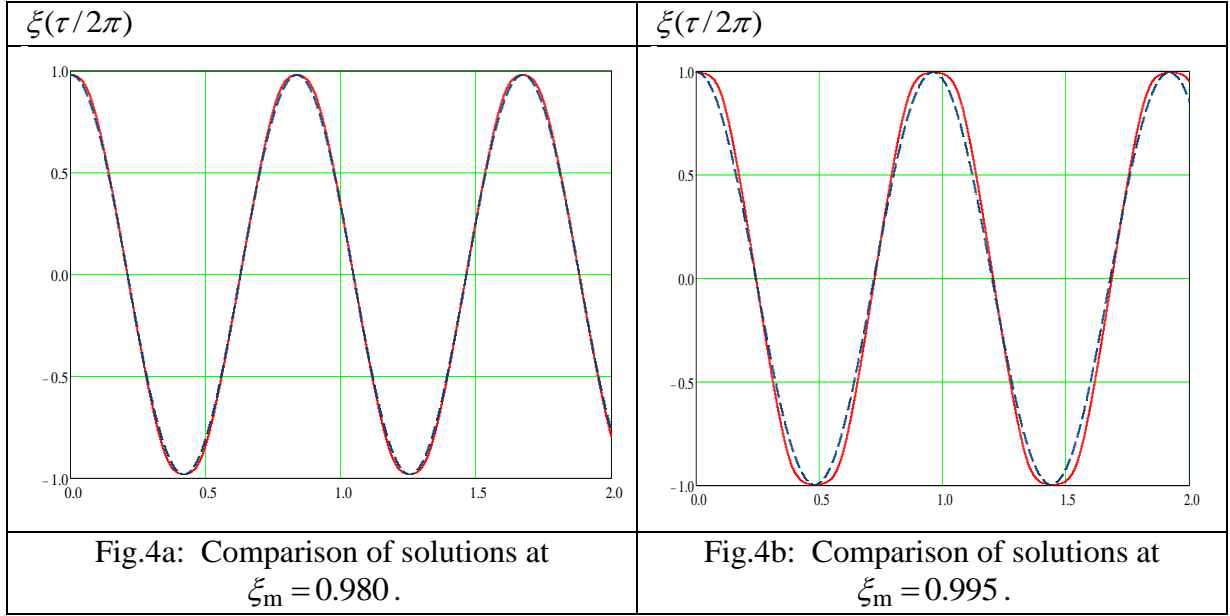

| Fig.4a: Comparison of solutions at $\xi_m = 0.980$. | Fig.4b: Comparison of solutions at $\xi_m = 0.995$. |

If periodic motion of a positron is considered in harmonic approximation (5) the expression for multipole parameter simplifies to

$$\rho(\xi_m) = 2\gamma^2\kappa^2 \langle p^2 \rangle = (\gamma\kappa\Omega(\xi_m)\xi_m)^2 \ .$$

Fig. 3 shows the dependence of multipole parameter (dashed line) calculated under this formula. It is seen that the harmonic approximation of periodic motion of positrons with energy of $E = 120$ GeV is quite acceptable for calculation of the radiation spectrum of channeled particles.

## §2. Distribution of density of channeled particles on amplitudes of motion

Besides dependence of frequency on amplitude of periodic motion for determining the full spectrum of radiation from all captured in the channeling positrons it is necessary to know:
- $N$     - a relative part of particles of the beam, captured into the channeling;
- $f(\xi_m)$ - density distribution of channeled positrons on amplitudes $\xi_m$.

We suppose that at the entry to the straight crystal positrons are distributed uniformly along the transverse coordinate $x$, and hence along $\xi$, and with the angular distribution $g(\vartheta)$. In normalized variables $(\xi, p)$ according to (2) we have the following relation between $\vartheta$ and $p$

$$p = \frac{d\xi}{d\tau} = \frac{\upsilon_s}{\omega_o(d/2)}\frac{dx}{ds} \cong \frac{2c}{\omega_o d}\vartheta = \frac{\vartheta}{\kappa} \ , \qquad (6)$$

where $s = \upsilon_s t \cong ct$ is the longitudinal coordinate along the channel and $1/\kappa \cong 6.230 \times 10^4$. From here the distribution of particles at the entry to the crystal on variable $p$ becomes $\tilde{g}(p) = \kappa g(\kappa p)$.

Closed phase curve $p = p(\xi, \xi_m)$ in the plane $\{p, \xi\}$ with a fixed $\xi_m$ for a channeled particle (see Fig. 5) according to (3, 4) is given by expression

$$p(\xi, \xi_m) = \pm\sqrt{2(\widetilde{U}(\xi_m) - \widetilde{U}(\xi))} \quad \text{with} \quad \xi \in [-\xi_m, \xi_m] \text{ and } 0 \leq \xi_m \leq 1. \tag{7}$$

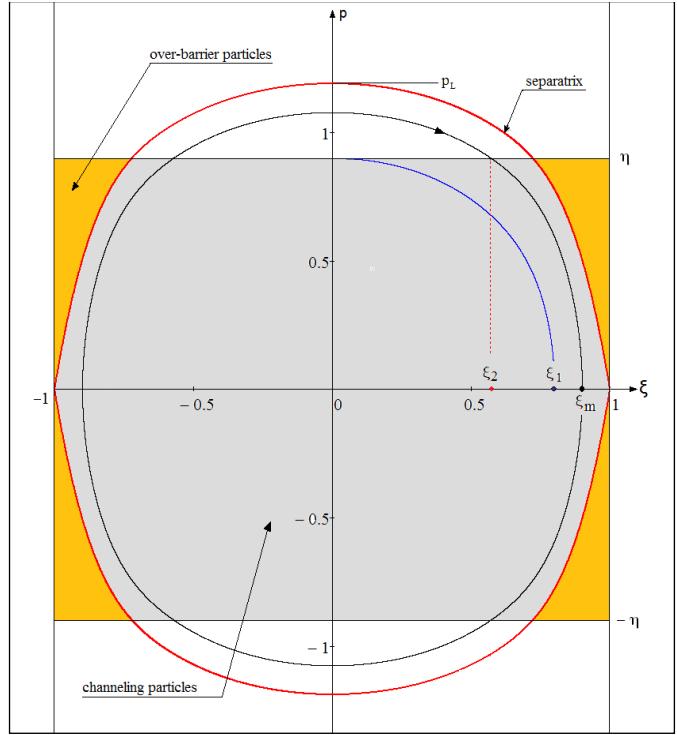

Fig.5: Phase portrait of capture of particles in the channeling.

Separatrix is the phase curve, separating the channeled and over-barrier particles, described by expression $p_c(\xi) = p(\xi, 1)$. The maximum value of $p_L$ corresponds to the Lindhard angle $\vartheta_L$, achieved at $\xi = 0$ and equal to $p_L = |p_c(0)| = \sqrt{2\widetilde{U}(1)}$. At the considered parameters of the straight crystal and the magnitude of the positrons energy we have according to (6): $p_L \cong 1.189$ and $\vartheta_L \cong 19.093 \times 10^{-6}$. Thus, the portion of particles captured in the channeling mode, i.e. moving inside the separatrix, is given by

$$N = \int_0^1 d\xi \int_{-|p_c(\xi)|}^{|p_c(\xi)|} \widetilde{g}(p)\,dp \;. \tag{8}$$

We are going now to find the density function $f(\xi_m)$ of particles distribution on the amplitudes only for the particles occurring in channeling. Hereinafter we mean $p(\xi, \xi_m)$ to be a positive branch of the definition (7). The relative number of channeled particles $N$ with amplitude $\leq \xi_m$ is equal to

$$F(\xi_m) = \frac{1}{N} \int_0^1 d\xi \int_{-p(\xi,\xi_m)}^{p(\xi,\xi_m)} \widetilde{g}(p)\,dp, \quad \text{i.e.} \quad F(1) = 1, \tag{9}$$

Then for the density function we get the expression

$$f(\xi_m) \equiv \frac{dF(\xi_m)}{d\xi_m} = \frac{1}{N}\frac{d\tilde{U}(\xi_m)}{d\xi_m}\int_0^{\xi_m}d\xi\,\frac{\tilde{g}(p(\xi,\xi_m))+\tilde{g}(-p(\xi,\xi_m))}{p(\xi,\xi_m)} \ . \qquad (10)$$

In the future, we confine ourselves to the simplest case of uniform and symmetric about zero angle distribution of particles, i.e.

$g(\vartheta) = \frac{1}{2\vartheta_o}\begin{cases}1 & \text{if } \vartheta \in [-\vartheta_o,\vartheta_o] \\ 0 & \text{if } \vartheta \notin [-\vartheta_o,\vartheta_o]\end{cases}$. Hence, by (6) in the plane of normalized variables we have $\tilde{g}(p) = \frac{1}{2\eta}\begin{cases}1 & \text{if } p \in [-\eta,\eta] \\ 0 & \text{if } p \notin [-\eta,\eta]\end{cases}$, here the boundary of the beam with variable $p$ is

$$\eta = \vartheta_o/\kappa.$$

**a)** For the case when the half-width of angular divergence more than the Lindhard angle, i.e. $\vartheta_o > \vartheta_L$ and, consequently, $\eta > p_L$, according to the preceding, we have:

$$N = \frac{1}{\eta}\int_0^1 |p_c(\xi)|d\xi = \frac{1}{\eta}\int_0^1 d\xi\sqrt{2(\tilde{U}(1)-\tilde{U}(\xi))} \ ,$$

$$F(\xi_m) = \frac{1}{N\eta}\int_0^{\xi_m}d\xi\, p(\xi,\xi_m) = \frac{1}{N\eta}\int_0^{\xi_m}d\xi\sqrt{2(\tilde{U}(\xi_m)-\tilde{U}(\xi))} \ ,$$

$$f(\xi_m) = \frac{1}{N\eta}\frac{d\tilde{U}(\xi_m)}{d\xi_m}\int_0^{\xi_m}\frac{d\xi}{\sqrt{2(\tilde{U}(\xi_m)-\tilde{U}(\xi))}} \ .$$

Thus, at $\vartheta_o > \vartheta_L$ the functions $F(\xi_m)$ and $f(\xi_m)$ do not depend on $\vartheta_o$.

Consider the approximation of a parabolic potential, where the above formulas are integrated until the end in the analytical form. For this we will limit ourselves in the representation of the potential $\tilde{U}(\xi)$ only by the first term of the expansion, i.e. $\tilde{U}(\xi) = \xi^2/2$. Then $\sqrt{2(\tilde{U}(\xi_m)-\tilde{U}(\xi))} \to \sqrt{2(\xi_m^2-\xi^2)}$. Hence:

$$N = \frac{\pi}{4\eta}\ , \qquad F(\xi_m) = \xi_m^2\ , \qquad f(\xi_m) = 2\xi_m \ .$$

**b)** For the case when the half-width of the angular spread of the beam is less than the Lindhard angle, $\vartheta_o < \vartheta_L$ and, hence, $\eta < p_L$ (see Fig. 5), we introduce the amplitude $\xi_1$, for which the phase curve has a maximum $p(0,\xi_1) = \eta$, i.e. it is determined from the equation $\sqrt{2\tilde{U}(\xi_1)} = \eta$. In addition, for every phase curve with the amplitude $\xi_1 < \xi_m \leq 1$ we determine the value $\xi_2$ which depends on $\xi_m$. Dependence $\xi_2(\xi_m)$ is determined by the equation $\sqrt{2(\tilde{U}(\xi_m)-\tilde{U}(\xi_2))} = \eta$. Note that $\xi_2(\xi_1) = 0$. According to the formulas of general form (8 ÷ 10) for our case of the angular spread in the beam of positrons, we get:

$$N = \xi_2(1) + \frac{1}{\eta}\int_{\xi_2(1)}^1 d\xi\sqrt{2(\tilde{U}(1)-\tilde{U}(\xi))} \ ; \qquad (11)$$

$$F(\xi_m) = \frac{1}{N} \begin{cases} \frac{1}{\eta} \cdot \int_0^{\xi_m} d\xi \sqrt{2(\tilde{U}(\xi_m) - \tilde{U}(\xi))} & \text{if } 0 \le \xi_m \le \xi_1 \\ \frac{1}{\eta} \cdot \int_{\xi_2(\xi_m)}^{\xi_m} d\xi \sqrt{2(\tilde{U}(\xi_m) - \tilde{U}(\xi))} + \xi_2(\xi_m) & \text{if } \xi_1 \le \xi_m \le 1 \end{cases},$$

$$f(\xi_m) = \frac{1}{N\eta} \frac{d\tilde{U}(\xi_m)}{d\xi_m} \begin{cases} \int_0^{\xi_m} \frac{d\xi}{\sqrt{2(\tilde{U}(\xi_m) - \tilde{U}(\xi))}} & \text{if } 0 \le \xi_m \le \xi_1 \\ \int_{\xi_2(\xi_m)}^{\xi_m} \frac{d\xi}{\sqrt{2(\tilde{U}(\xi_m) - \tilde{U}(\xi))}} & \text{if } \xi_1 \le \xi_m \le 1 \end{cases}. \quad (12)$$

Thus, at $\vartheta_o < \vartheta_L$ in a parabolic approximation, we obtain $\xi_1 = \eta$ and $\xi_2(\xi_m) = \sqrt{\xi_m^2 - \eta^2}$, and the above formula reduces to

$$N = \frac{1}{2}\sqrt{1-\eta^2} + \frac{1}{2\eta}\arcsin(\eta)$$

$$F(\xi_m) = \frac{1}{2N\eta} \begin{cases} \pi \xi_m^2 / 2 & \text{if } 0 \le \xi_m \le \eta \\ \eta\sqrt{1-\eta^2} + \xi_m^2 \arcsin(\eta/\xi_m) & \text{if } \eta \le \xi_m \le 1 \end{cases}$$

$$f(\xi_m) = \frac{\xi_m}{N\eta} \begin{cases} \pi/2 & \text{if } 0 \le \xi_m \le \eta \\ \arcsin(\eta/\xi_m) & \text{if } \eta \le \xi_m \le 1 \end{cases}.$$

Calculated by the expression (11) dependence of the relative capture $N$ in the channeling regime on the value of the half-width of the angular spread $\vartheta_o$ of the beam for the potential (1) is shown in Fig. 6. Calculated according to (12) density functions $f(\xi_m)$ for some values $\vartheta_o \times 10^6$ are shown in Fig. 7.

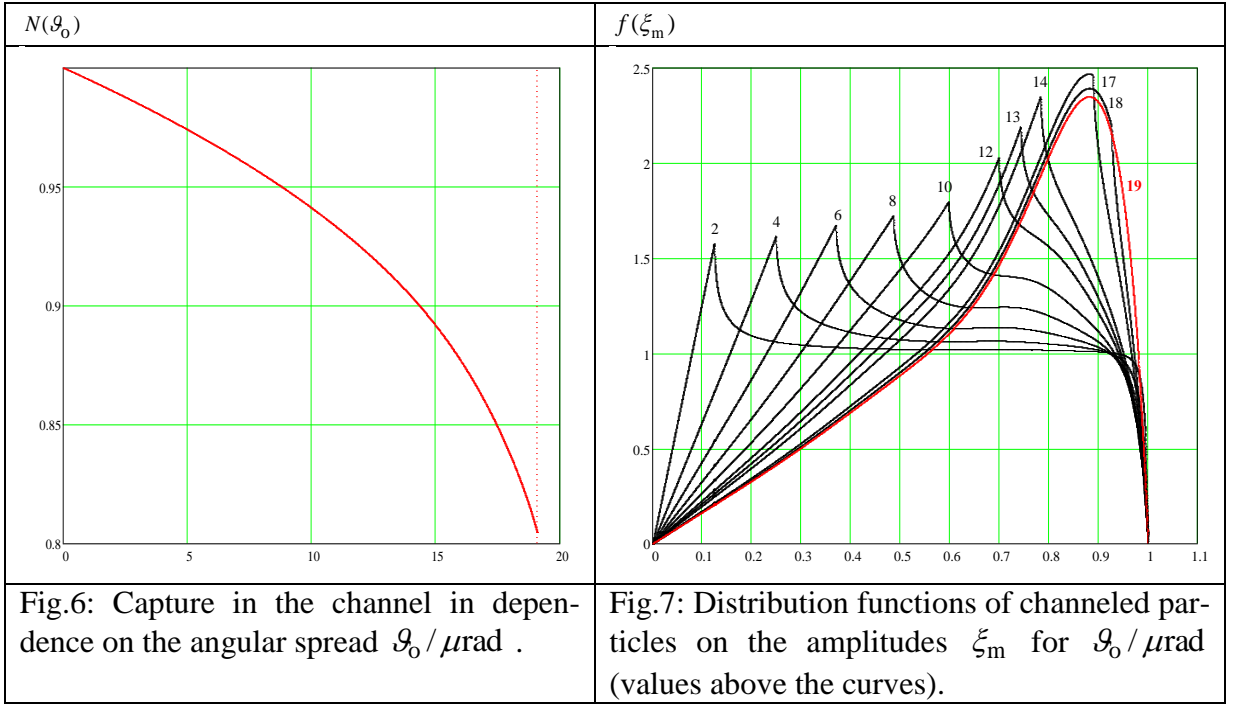

Fig.6: Capture in the channel in dependence on the angular spread $\vartheta_o / \mu$rad.

Fig.7: Distribution functions of channeled particles on the amplitudes $\xi_m$ for $\vartheta_o / \mu$rad (values above the curves).

From the above analysis it follows the further important conclusion: in the potential (1), where $d\tilde{U}/d\xi \to 0$ at $\xi \to \pm 1$ the density distribution $f(\xi_m) \to 0$ at $\xi_m \to 1$. Thus, a small (almost zero) density of the channeled particles is shown for oscillations with amplitudes close to unity. This once again makes the approximation of the harmonic nature of the motion of channeled positrons in a crystal true enough.

## §3. The radiation of channeled positrons in quasiperiodic motion

To find the radiation spectrum of channeled positrons, oscillating in the interplanar potential (1), use the formula derived in [1] (p.303) for the quasiperiodic motion of a particle at all values of multipole parameter $\rho$, taking into account the quantum recoil effect after the emission of a photon. This formula is valid for calculations of systems in which the particle performs sufficiently large (ten or more) number of oscillations along a straight line. In deriving the formula the coupling of the transverse and longitudinal motions was taken into account. The need to consider the radiation spectrum in such a very general way is due to the fact that in the potential (1) multipole parameter (see Fig. 3) covers a wide range of values $\rho$, providing different types of radiation (see Introduction). The radiation spectrum of one positron per unit length of a short crystal is determined by the following expression:

$$\frac{d^2 E}{dE_\gamma ds} = -\frac{\alpha E_\gamma}{c(2\pi\gamma)^2} \sum_{n=1}^{\infty} \Phi(n - \zeta \cdot (1+\rho/2)) \int_{-\pi}^{\pi} \int_{-\pi}^{\pi} d\varphi_1 d\varphi_2 \, J_0\left(2 \int_{\varphi_2}^{\varphi_1} d\varphi \mu(\varphi) \sqrt{\zeta \cdot (n - \zeta \cdot (1+\rho/2))}\right) \times$$

$$\times \left[1 + \frac{A(E_\gamma)}{2}(\mu(\varphi_1) - \mu(\varphi_2))^2\right] \times \cos\left\{(n - \zeta \cdot \rho/2)(\varphi_1 - \varphi_2) + \zeta \int_{\varphi_2}^{\varphi_1} d\varphi \mu^2(\varphi)\right\} \quad (13)$$

where: $\alpha = 1/137.04$, $E_\gamma$ is the energy of the emitted photon,

step function $\Phi(y) = 1$ at $y \geq 0$ and $= 0$ at $y < 0$,

$J_0$ is the Bessel function, $\mu(\varphi) = \gamma(\upsilon_x(\varphi) - <\upsilon_x>)/c$ (in the potential (1) for the channeled positron $<\upsilon_x> = 0$),

$$\zeta(E_\gamma, \xi_m) = \frac{E_\gamma E}{2\gamma^2(\hbar\omega)(E - E_\gamma)}, \quad \omega = \omega_0 \cdot \Omega(\xi_m), \quad A(E_\gamma) = 1 + \frac{E_\gamma^2}{2E(E - E_\gamma)}.$$

According to the previous analysis the motion of a positron in the potential well (1) is presented in the form of a harmonic oscillation (5) with the frequency depending on its amplitude $\xi_m$. For such an approximation the spectral dependency $(d^2 E/dE_\gamma ds)$ on $E_\gamma$ was calculated by the previous formula and shown in Fig. 8 for a single channeled positron with the following values of $\xi_m = $ 0.3, 0.5, 0.7, 0.9.

In calculating the spectrum for each value of $E_\gamma$ the summation over $n$ in (13) ended at $n_{max}(E_\gamma)$, when the addition of the following term with $n = n_{max}(E_\gamma) + 1$ in relation to the accumulated sum became less than 1%. Integration was carried out with precision $10^{-3}$. Full spectral emission of all positrons trapped in the channel is given by

$$\left(\frac{d^2E}{dE_\gamma ds}\right)_{tot} = N \int_0^1 f(\xi_m)\Psi(E_\gamma,\xi_m)\,d\xi_m \quad, \tag{14}$$

where $\Psi(E_\gamma,\xi_m)$ is a function of spectral intensity of one positron with amplitude $\xi_m$, i.e. right-hand side of the expression (13). Calculated by formula (14) dependence $(d^2E/dE_\gamma ds)_{tot}$ on $E_\gamma$ for two cases of the initial half-width of the angular spread of the beam of positrons $\vartheta_o \times 10^6 = 4$ and $\vartheta_o \times 10^6 = 19$ is shown in Fig. 9. The maximum level of the full spectrum of radiation and the corresponding value of the photon energy $E_\gamma$ are shown in the following table.

| $\vartheta_o \times 10^6$ | $\eta$ | N | $\max(d^2E/dE_\gamma ds)_{tot}$ | $E_\gamma$/GeV |
|---|---|---|---|---|
| 4 | 0.249 | 97.96 % | ≈4.2/см | ≈(1.95÷2.30) |
| 19 | 1.184 | 80.79 % | ≈5.1/см | ≈(2.27÷2.96) |

As noted above, formula (13) describes the radiation for any value of the multipole parameter $\rho$, including the quantum domain when you need to take into account the recoil after the emission of a photon. If we ignore this recoil, the functions $\zeta(E_\gamma,\xi_m)$ and $A(E_\gamma)$ in (13) take the form

$$\zeta(E_\gamma,\xi_m) = \frac{E_\gamma}{2\gamma^2(\hbar\omega)}, \qquad A(E_\gamma) = 1 + \frac{E_\gamma^2}{2E^2}.$$

For this case Fig. 12 shows the spectral dependence (pointed line **b**) for all channeled positrons at $\vartheta_o \times 10^6 = 19$.

## §4. The calculation on the basis of relations for magnetic bremsstrahlung radiation

Besides of calculating the radiation by use of method [1] (formula (1)), we consider the process of radiation in the channeling of positrons by another method, based on the equations of classical electrodynamics. This method is valid when the direction of motion of a particle moving in a field is changing at an angle substantially higher than the characteristic angle of radiation $\sim 1/\gamma$. The essence of the method consists in the fact that we only consider the transverse motion and neglect the interference from distant parts of the trajectory, as it is realized in the case of magnetic bremsstrahlung radiation. Then according to [15] (see p.258) the spectral radiation with frequency of photons $\tilde{\omega}$ (or energy $E_\gamma = \hbar\tilde{\omega}$) per unit length $s$, generated by ultrarelativistic channeled positron, performing harmonic transverse oscillations with a frequency $\omega(\xi_m)$, is determined by the expression:

$$\frac{d^2E}{dE_\gamma ds} = -M(\xi_m)\frac{2e^2}{\gamma^2 c\hbar} n\, 4 \int_0^{\pi/2}\left\{\frac{\text{Ai}'(u)}{u} + \frac{1}{2}\int_u^\infty \text{Ai}(s)\,ds\right\}d\varphi \quad, \tag{15}$$

where: $E_\gamma = \hbar \widetilde{\omega}$, $\hbar$ – Planck constant, $M(\xi_m) = \dfrac{\omega(\xi_m)}{2\pi c}$ - number of oscillations of a particle passing a unit length of the crystal, $n = \dfrac{\widetilde{\omega}}{\omega(\xi_m)}$, $\mathrm{Ai}(u) = \dfrac{1}{\pi} \int_0^\infty \cos\left(\dfrac{y^3}{3} + u\, y\right) dy$ - Airy function, $\mathrm{Ai}'(u) = \dfrac{d}{du}\mathrm{Ai}(u)$, $u = u(\xi_m, \varphi) = \chi(\xi_m) \cdot \left(\dfrac{n}{\cos(\varphi)}\right)^{\tfrac{2}{3}}$, $\chi(\xi_m) = \dfrac{1}{\gamma^2}\left(\dfrac{c}{\omega(\xi_m)}\dfrac{2}{\xi_m d}\right)^{\tfrac{2}{3}}$.

As before, the motion of a positron in a potential well (1) is represented as a harmonic oscillation (5) with frequency depending on amplitude $\xi_m$. For such an approximation the spectral dependences $(d^2E/dE_\gamma ds)$ on $E_\gamma$ were calculated by the previous formula (15) and shown in Fig. 10 (similar to Fig. 8) for a single channeled positron with the following values of the amplitudes $\xi_m = 0.3, 0.5, 0.7, 0.9, 0.940, 0.995, 0.999$. So now the total radiation of all particles captured in the channel is given by the expression

$$\left(\dfrac{d^2 E}{dE_\gamma ds}\right)_{tot} = N \int_0^1 f(\xi_m) L(E_\gamma, \xi_m)\, d\xi_m, \qquad (16)$$

where $L(E_\gamma, \xi_m)$ is a function of the spectral intensity of a positron with amplitude $\xi_m$, i.e. right-hand side of expression (15). The dependence $(d^2E/dE_\gamma ds)_{tot}$ on $E_\gamma$ calculated by formula (16) for two cases of the initial half-width of the angular spread of the positron beam $\vartheta_o/\mu rad = 4$ and $\vartheta_o/\mu rad = 19$ is shown in Fig. 11.

Fig. 12 shows the spectral curves obtained for the case of $\vartheta_o/\mu rad = 19$ using two methods from [1] taking into account the recoil (curve **a**) and without it (curve **b**) and from [15] using classical electrodynamics (curve **c**). Comparison of these dependences shows their satisfactory agreement: the maximum of the radiation spectrum obtained by means of the classical electrodynamics exceeds such a maximum taking into account the recoil less than 10%.

## Discussion of results

The planar periodic motion of a positron channeled in straight crystal with orientation (011) was discussed in the paper in some detail: dependences of the oscillation frequency and of the multipole parameter on the amplitudes have been found. A wide range of values of the multipole parameter in the considered potential points out the realization of various types of radiation: interference – mostly by the particles with small amplitudes and magnetic bremsstrahlung - with relatively large amplitudes. It is shown that in such a potential the density distribution of channeled particles on the amplitude $\xi_m$ tends to zero at $\xi_m \to 1$. These are the particles which, while moving, approach close to the crystallographic planes. The portion of such particles is small and, furthermore, these particles are more likely to leave the channeling regime, i.e. be dechanneled due to interaction with nuclei of the crystal lattice.

It is shown that for the main part of the channeled particles with amplitudes less than ≈0.98, their motion is the harmonic oscillations with calculated frequencies, depending on their amplitudes. The calculation of the radiation spectrum of channeled particles in this harmonic approximation is greatly simplified in two ways: the method used in [1] with and without taking into account the quantum recoil and the method of classical electrodynamics [15]. Comparison of the results shows their satisfactory agreement: the maximum of the radiation spectrum received by means of the classical electrodynamics exceeds the maximum received with taking into account the recoil by less than 10%.

Thus, because the method of [1] applies only to a straight crystal, the method of classical electrodynamics [15] may serve as a basis for further consideration of the radiation process in a bent monocrystal.

# Conclusion

The calculations prove the prospects for the application of monocrystals to create powerful sources of directed radiation at the accelerators and will be used for comparison with experimental data INSURAD and when planning new studies at CERN and SRC IHEP.

Work was supported by the Direction SRC IHEP and RFFI grants №№ 07-02-00022-a, 08-02-01453-a and also grant № 09-02-92431-КЭ_a of the joint project RFFI – Consortium EINSTEIN (Italy).

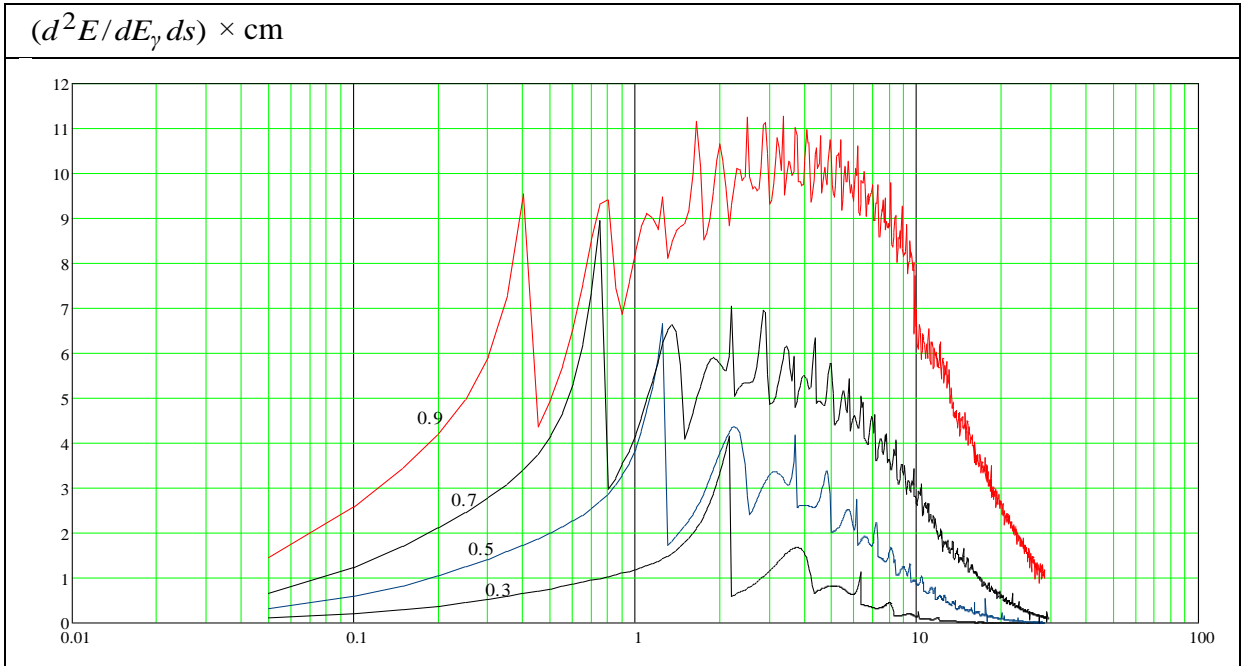

Fig.8: Spectral dependence $(d^2E/dE_\gamma ds)$ on $E_\gamma$/GeV, obtained by method of [1] taking into account the recoil during the radiation for the values $\xi_m$ marked on each curve.

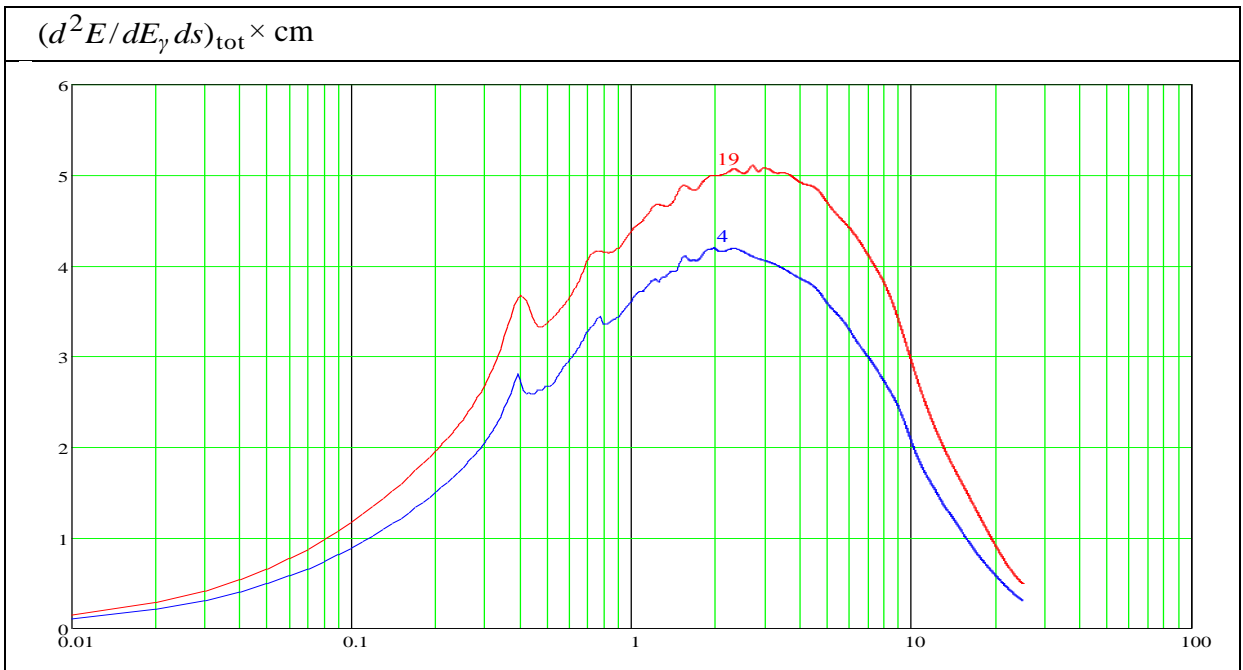

Fig.9: Total spectral intensity depending on the energy of the emitted photons $E_\gamma$/GeV obtained by the method of [1] taking into account the recoil during the radiation.

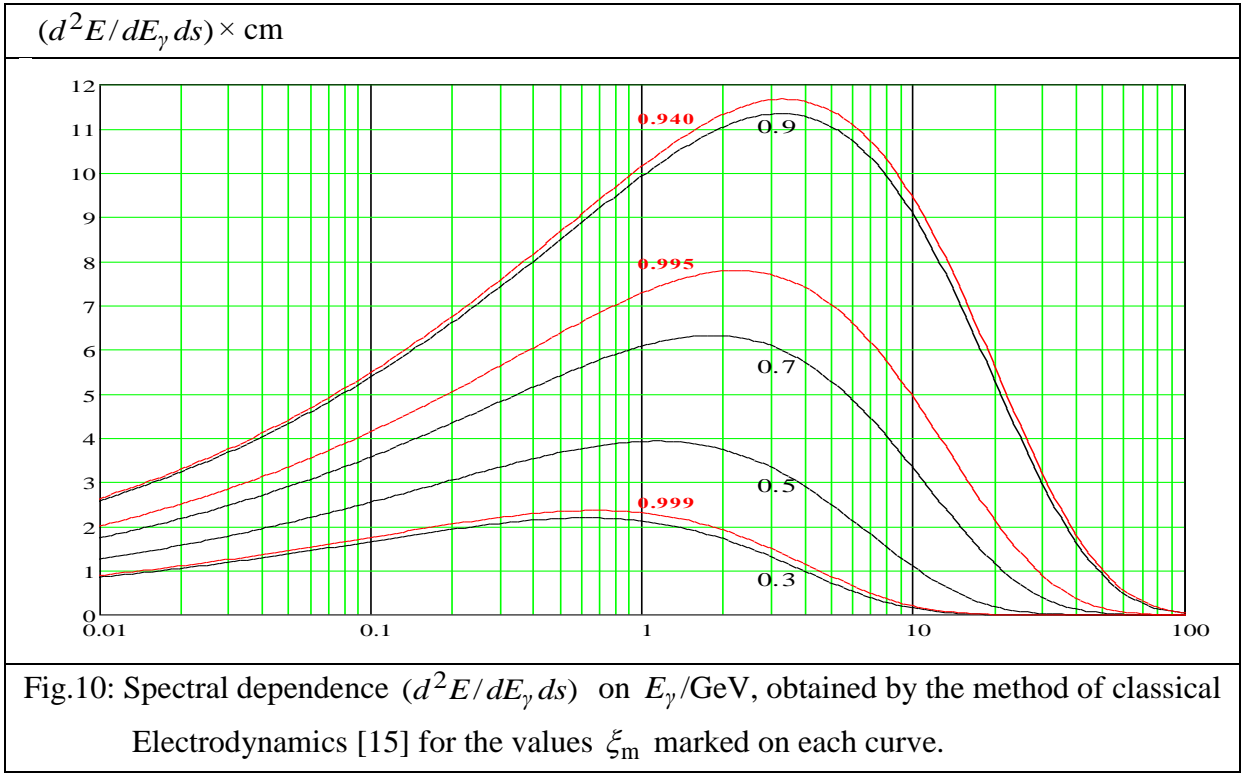

Fig.10: Spectral dependence $(d^2E/dE_\gamma ds)$ on $E_\gamma$/GeV, obtained by the method of classical Electrodynamics [15] for the values $\xi_m$ marked on each curve.

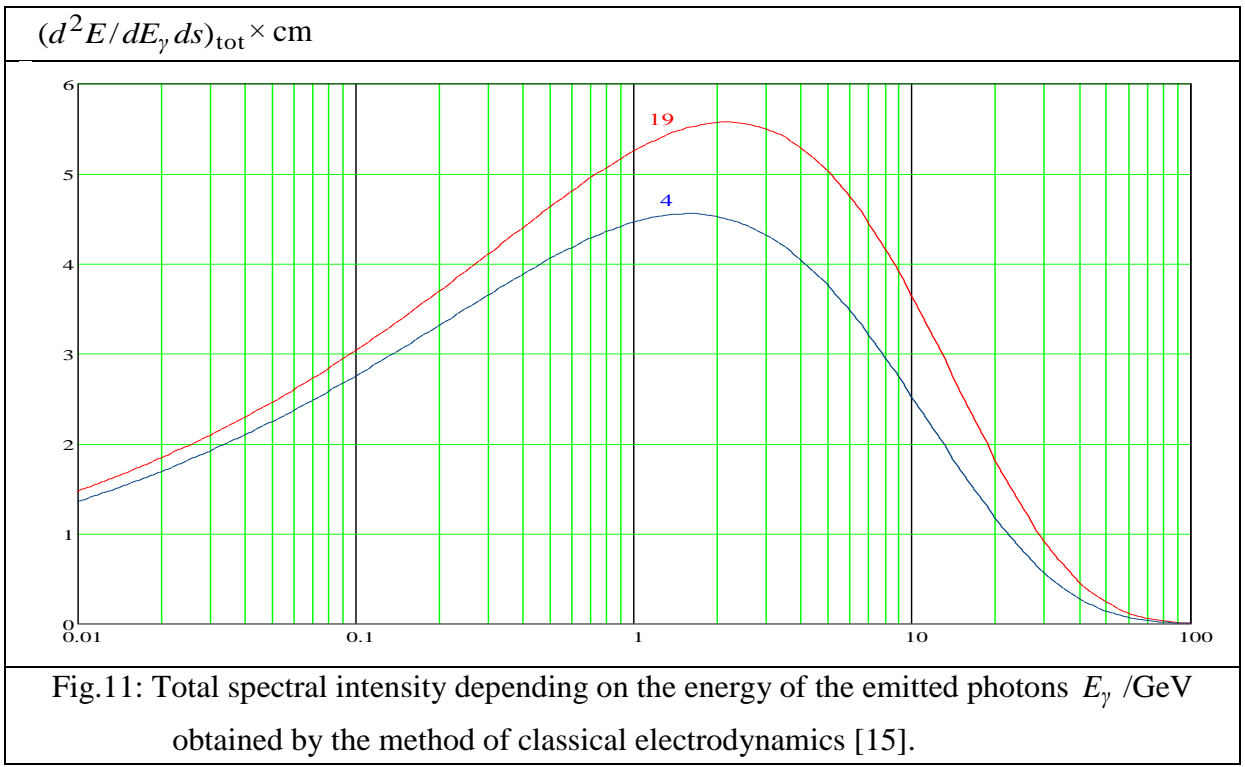

Fig.11: Total spectral intensity depending on the energy of the emitted photons $E_\gamma$/GeV obtained by the method of classical electrodynamics [15].

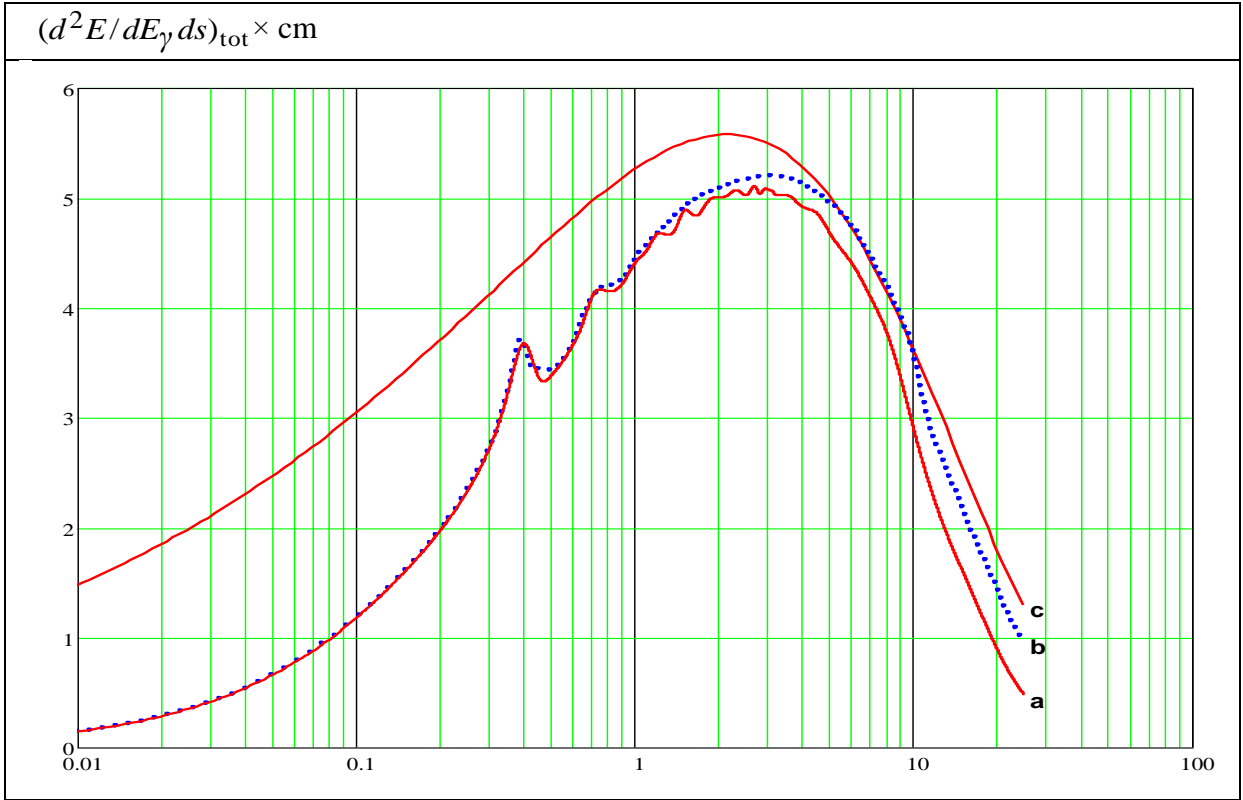

Fig.12: The total spectral intensity depending on the energy of the emitted photons $E_\gamma$ /GeV at $\vartheta_o = 19\,\mu$rad calculated by the method of [1] taking into account the recoil (curve **a**) and without it (curve **b**), method of classical electrodynamics [15] (curve **c**).